 \documentclass[final,5p,times,twocolumn]{elsarticle}

 \usepackage{graphics}

\usepackage{amssymb}
\usepackage{amsmath}

\journal{Physica E}

\begin{document}

\begin{frontmatter}


\title{Interplay of intrinsic and extrinsic mechanisms to the spin\\ Hall effect in a  two-dimensional electron gas} 


\author[address1]{Roberto Raimondi\corref{cor1}}
\author[address2]{Peter Schwab}

\address[address1]{Dipartimento di Fisica "E. Amaldi", Universit\`a  Roma Tre, Via della Vasca Navale 84, 00146 Roma, Italy}
\address[address2]{Institut f\"ur Physik, Universit\"at Augsburg, 86135 Augsburg, Germany}

\cortext[cor1]{
Corresponding author.
E-mail: raimondi@fis.uniroma3.it}

\begin{abstract}
In order to describe correctly the interplay of extrinsic and intrinsic spin-orbit mechanisms 
to the spin Hall effect, it is necessary to consider different
sources of spin relaxation. We take into account the spin relaxation
time $\tau_{DP}$ due
to the Dyakonov-Perel mechanism as well as the Elliot-Yafet 
spin-relaxation time $\tau_s$ due to the spin-orbit scattering from impurities. 
The total spin Hall conductivity depends crucially  
on the ratio $\tau_s /\tau_{DP}$.
\end{abstract}


\end{frontmatter}


\section{Introduction}
In a recent paper\cite{raimondi2009}, we have provided a general
framework to describe 
the electric-field control of the electron spin in a two-dimensional
electron gas (2DEG) with diffusive electron spin dynamics when both intrinsic and extrinsic spin-orbit
interactions are present. 
The extrinsic mechanism for the spin Hall effect arises to first order in the spin-orbit coupling strength, 
$\lambda_0^2$ (see below)\cite{engel2005,tse2006}. 
However, when also the intrinsic mechanism is present\cite{tse2006a,hu2006,hankiewicz2008}, 
a first-order calculation is no longer sufficient, since it is important to consider 
the Elliot-Yafet spin relaxation due to the spin-orbit scattering from impurities, 
which arises in the order  $\lambda_0^4$.
Here, by using a Keldysh Green function approach\cite{raimondi2006}, 
we provide a  microscopic basis of the equation of motion for the spin
density that governs the interplay of intrinsic and extrinsic mechanisms.
The layout of the paper is as follows. In the next section we introduce the model Hamiltonian
and the method. In section 3 we derive an expression for the spin current and an associated continuity equation.  In section 4, we consider the specific case of the spin Hall effect and derive a formula for the spin Hall conductivity. Finally, we state our conclusions in section 5. 

\section{The model and the method}
In the presence of both extrinsic and intrinsic spin-orbit interaction as well as normal potential scattering from impurities $V({\bf x})$, 
the Hamiltonian for the 2DEG can be written in terms of 
a spin-dependent [$SU(2)$] vector potential ${\bf \tilde A}$
\begin{equation}
\label{hamiltonian}
H=\frac{{\bf p}^2}{2m}-\frac{{\bf \tilde A}\cdot {\bf p}}{m}
  + V({\bf x}).
\end{equation}
The vector potential is the sum of  intrinsic (Rashba type) and extrinsic contributions
\begin{eqnarray}
  {\bf \tilde A}  & = & m \ {\boldsymbol \sigma}\times \left(\alpha { \bf \hat e}_z +\frac{\lambda_0^2}{4}
  \partial_{\bf x} V({\bf x})\right)
        \equiv \frac{1}{2}{\bf \tilde A}^a\sigma^a, \label{vectorpotential}
\end{eqnarray}
where ${\boldsymbol \sigma}$ is the vector of Pauli matrices and $ a=x,y,z$.
The coupling constants $\alpha$ and $\lambda_0^2$ characterize the
strength of the intrinsic and extrinsic spin-orbit interaction, respectively. 
The advantage of introducing the $SU(2)$ vector potential is that one can immediately derive a 
continuity equation for the spin density 
\begin{equation}
\label{operatorform}
\partial_t s^a +{\partial}_{\bf x}\cdot  {\bf j}^a 
 + \varepsilon_{abc} {\tilde {\bf A}}^b \cdot {\bf j}^c = 0,
\end{equation}
where the spin density is defined in terms of the Heisenberg field operators
$s^a=(1/2) \langle  \psi^{\dagger}({\bf x}, t) \sigma^a \psi({\bf x}, t) \rangle $ 
and the spin current has the expression 
${\bf j}^a=(1/4) \langle \psi^{\dagger}({\bf x}, t)\lbrace \sigma^a , {\bf v}\rbrace \psi({\bf x}, t) \rangle$,
with the velocity operator given by
\begin{equation}
\label{velocity}
{\bf v}=\frac{{\bf p}-{\bf \tilde A}}{m}.
\end{equation}
Although Eq.~(\ref{operatorform}) is formally exact, it cannot directly convey information for the
{\it disorder averaged} spin current, since the disorder potential appears also in 
the vector potential $\bf \tilde A$. In order to carry out the average over the disorder, 
it is convenient to use the Green function approach as developed in
Ref.~\cite{raimondi2006}. 
It is also useful to explicitly separate the spin-orbit interaction due to the intrinsic mechanism from 
that due to the extrinsic mechanism. 
To this end we define a space-independent vector potential 
\begin{equation}
{\bf A}  =  \alpha m \ {\boldsymbol \sigma}\times  { \bf \hat e}_z  \label{rashbavectorpotential}
\end{equation}
and a spin-dependent disorder potential
\begin{equation}
U({\bf x})  =  V({\bf x})- \frac{\lambda_0^2}{4}{\boldsymbol \sigma}\times
 \partial_{\bf x} V({\bf x})\cdot {\bf p}.\label{effectivepotential}
\end{equation}
The {\em disorder averaged}  spin density  is given by
\begin{equation}
\label{observables}
s^a({\bf x}, t)=
-\frac{{\rm i}}{2} 
  \int \frac{d \epsilon}{2 \pi} \int \frac{d^2 p}{(2\pi)^2} 
  \mathrm{Tr}(\sigma^a { G^<}(\epsilon, {\bf p},{\bf x}, t))
, \end{equation} 
where  ${ G^<}(\epsilon, {\bf p},{\bf x}, t)$ is the lesser component
of the disorder averaged Green function, here given in Wigner coordinates\cite{rammer1986}.
The current density is written as a sum of two terms, ${\bf j}^a={\bf j}_0^a+{\bf j}_{av}^a$, associated to the normal,
${\bf v}_0=({\bf p}-{\bf A})/m$, and to the anomalous disorder-dependent components of the
velocity, respectively. Explicitly  
the normal current is
\begin{equation}
\label{observablescurrent}
{\bf j}^a_0({\bf x}, t)=
-\frac{{\rm i}}{4}
  \int \frac{d \epsilon}{2 \pi} \int \frac{d^2 p}{(2\pi)^2}
    \mathrm{Tr}[ \sigma^a \lbrace {\bf v}_0  ,  { G^<}(\epsilon, {\bf
    p},{\bf x}, t) \rbrace ]
    , \end{equation}
where the symbol $\lbrace, \rbrace$ denotes the anticommutator.
Since the anomalous velocity contribution to the current contains explicitly the disorder potential,
its expression can be obtained only after specifying the disorder model and the approximations
used. Later on, we will identify  the expression for the anomalous
current  from the self-energy terms contributing to the continuity equation.

In Wigner coordinates and after a gradient expansion, the equation of
motion for the Green function 
$\check{G}(\epsilon, \mathbf{ p},\mathbf{ x}, t )$ reads
\begin{equation}
\label{equationofmotion}
{\rm i}{\partial}_t {\check G}+
{\rm i} {\partial}_{\bf x}
   \cdot \Big\{ \frac{{\bf p}-{\bf A}}{2m}, {\check G} \Big\}
 +   \left[  \frac{{\bf A} \cdot {\bf p} }{m}, \check G  \right]
= \left[ {\check \Sigma }, {\check G}\right].
\end{equation}
The Green function has a matrix structure in both the Keldysh and spin
space, and the symbol
$[,]$ indicates a commutator.
An external electric field ${\bf E}$ is included  via the substitution
$\partial_{\bf x} \to \partial_{\bf x} - e {\bf E} \partial_\epsilon$.
The self-energy $\check\Sigma$  entering the RHS of the Eq.(\ref{equationofmotion}) contains all the effect of the
disorder potential defined in Eq.(\ref{effectivepotential}).
We assume the standard model of uncorrelated impurities with
\begin{eqnarray}
    \overline{ V({\bf x}_1)V({\bf x}_2) }  =  n_i v_o^2\delta ({\bf x}_1-{\bf x}_2),
\label{disordermodel}
\end{eqnarray} 
where $n_i$ is the impurity concentration and $v_0$ the scattering amplitude. 
\begin{figure}[t]
\begin{center}\leavevmode
\includegraphics[width=0.6\linewidth]{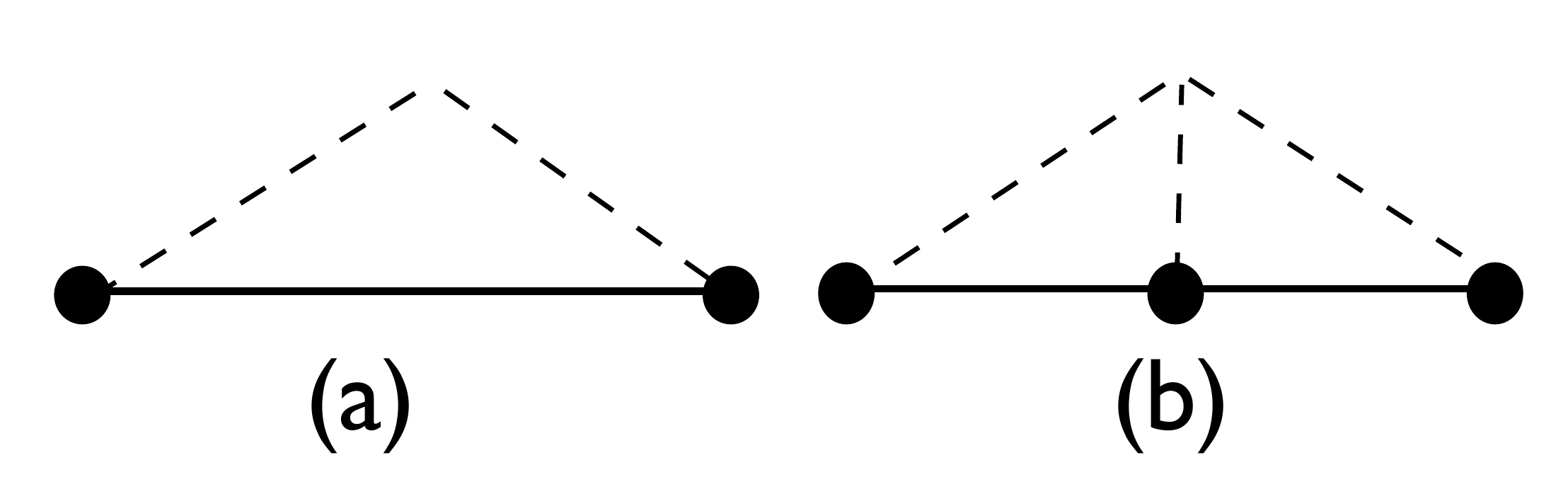}
\caption{Lowest order diagrams for the disorder averaged self-energy.
(a) Born approximation. (b) Third order contribution.
The full dot indicates the insertion of the potential $U({\bf x})$ (cf. Eq.(\ref{effectivepotential})), whereas the dashed line represents
the disorder average.}
 \label{selfenergy}
\end{center}
\end{figure}
In the Born approximation it is sufficient to define the moments of the disorder distribution up to the second order as done in Eq.(\ref{disordermodel}). At the level of the Born approximation, the self-energy is given by 
the diagram of Fig.~\ref{selfenergy}(a) and reads
\begin{equation}
 \label{born}
 {\check \Sigma}({\bf x}_1, {\bf x}_2)= \overline{ U({\bf x}_1) {\check G}({\bf x}_1, {\bf x}_2)U({\bf x}_2)}.
\end{equation}
To zero order in $\lambda_0^2$, Eq.~(\ref{born}) yields 
\begin{equation}
 \check\Sigma^0({\bf p}, {\bf x})=n_i v_o^2 \sum_{{\bf p}' } \check G ({\bf p}', {\bf x}) \label{scatteringtime}
.\end{equation}
This contribution leads to the standard  elastic scattering time
$\tau^{-1}=2\pi N_0 n_iv_o^2$ \cite{schwab2003}, where $N_0$ is the single-particle density of states
of the Fermi gas.
To first order in $\lambda_0^2$, Eq.~(\ref{born}) yields the three terms
\begin{eqnarray}
\check\Sigma^1_{a}({\bf p}, {\bf x} )  & = & - {\rm i} \frac{\lambda_0^2}{ 4} n_i v_0^2
  \epsilon_{abc} \sum_{{\bf p}' }  \left[{ p}_a  \sigma_b {p'_c} ,
   \check G({\bf p}', {\bf x}) \right], \\
\check\Sigma^1_{b}({\bf p}, {\bf x} )  & = & \frac{\lambda_0^2}{8}n_iv_o^2 
  \epsilon_{abc } \sum_{{\bf p}'}\lbrace  \sigma_b p'_c,\partial_a \check G({\bf {p}'},{\bf x})\rbrace 
 \label{sjcurrent},\\
\check\Sigma^1_{c}({\bf p}, {\bf x}) & = & \frac{\lambda_0^2}{8}n_iv_o^2 
 \epsilon_{abc} \sum_{{\bf p}'}\lbrace {p}_a \sigma_b ,
 \partial_{c} \check G({\bf p}',{\bf x})\rbrace
 \label{sjelectric}
; \end{eqnarray}
again the external electric field is included in the 
space derivative $\partial_a \equiv \partial_{x_a}-eE_a\partial_{\epsilon}$.
The self-energy $\check\Sigma^1_a$ is related to the mechanism for "swapping of spin
currents" \cite{lifshits2009}, $\check\Sigma^{1}_b$ and $\check\Sigma^1_c$ are
side-jump contributions. Notice that to lowest order in the gradient expansion, only the
"swapping term" is different from zero. The side-jump contributions arise when considering the next-to-leading order in the gradient expansion. 
Finally, to second order in $\lambda_0^2$, Eq.~(\ref{born}) yields 
\begin{equation}
\label{taus}
\check\Sigma^2  = \frac{\lambda_0^4}{16}n_iv_o^2\sum_{\bf k}\sigma^z\check G
({\bf k},{\bf x})\sigma^z ({\bf p}\times {\bf k})_z^2,
\end{equation}
which corresponds to the Elliot-Yafet spin relaxation mechanism.

Skew-scattering arises beyond the Born
approximation\cite{engel2005,tse2006}, starting from the third order
diagram shown in Fig.~\ref{selfenergy}(b).
By using a disorder model with third moments different from zero and defined by\cite{tse2006}
\begin{equation}
\overline{ V({\bf x}_1)V({\bf x}_2)V({\bf x}_3) }=n_iv_o^3\delta ({\bf x}_1-{\bf x}_2)\delta ({\bf x}_2-{\bf x}_3),
\end{equation}
the diagram of  Fig.~\ref{selfenergy}(b) yields 
to first order in $\lambda_0^2$, the terms
\begin{eqnarray}
\check\Sigma^1_{SS,a}&=& {\rm i}\frac{\lambda_0^2}{4}n_iv_o^3\sum_{{\bf k},{\bf k}'}
\check G ({\bf k}, {\bf x})\check G ({\bf k}', {\bf x}){\bf p}\times {\bf k}'\cdot {\boldsymbol \sigma}\label{ssa}\\
\check\Sigma^1_{SS,b}&=&  {\rm i}\frac{\lambda_0^2}{4}n_iv_o^3
\sum_{{\bf k},{\bf k}'}
{\bf k}\times {\bf p}\cdot {\boldsymbol \sigma}
\check G ({\bf k}, {\bf x})\check G ({\bf k}', {\bf x}).\label{ssb}
\end{eqnarray}
To make contact with the diagrammatic language of the Kubo formula, we notice that the two self-energies (\ref{ssa}-\ref{ssb}) correspond to the diagrams of Fig.2 in Ref.\cite{tse2006}.

\section{The continuity equation and the spin current}
The insertion of Eqs.~(\ref{scatteringtime}--\ref{taus}) and (\ref{ssa}--\ref{ssb}) into the
equation-of-motion (\ref{equationofmotion}) allows to derive a continuity equation for the spin density
polarized along the $a$-axis.
After integrating over ($\epsilon$,  ${\bf p}$) and taking the trace of Eq.~(\ref{equationofmotion}), one obtains
\begin{equation}
\label{continuity}
{\partial}_t s^a  + \tilde { \partial}_{\bf x}\cdot {\bf j}^a_0 =
- \partial_x \cdot {\bf j}^a_{av}  -  \frac{1}{\tau_s} s^a  
.\end{equation}
The gradient together with the commutator on the LHS of
Eq.~(\ref{equationofmotion}) are the origin of the covariant
derivative of the spin current
\begin{equation}
\label{covariant}
\tilde { \partial}_{\bf x}\cdot {\bf j}^a_0 = 
{ \partial}_{\bf
x}\cdot {\bf j}^a_0 + \varepsilon_{abc} {\bf A}^b \cdot {\bf j}^c_0 
.\end{equation}
Most of the contributions to the self-energy disappear
in the integrated equation. The only terms surviving the integration are those originating from
$\check\Sigma^1_b$ and $\check\Sigma^2$.
The term containing $\check\Sigma^1_b$ of Eq.~(\ref{sjcurrent}) yields a contribution which can be written as a divergence and hence, apart from a minus sign, defines the anomalous contribution to the current  ${\bf j}_{av}^a$. 
The spin relaxation time, which  is obtained from Eq.~(\ref{taus}), only applies to the in-plane spin components 
(cf. the two $\sigma^z$ matrices before and after the Green function) and reads
$\tau_s^{-1}=\tau^{-1}(\lambda_0p_F/2)^4$.

Clearly, also the explicit expression for the spin current can be
obtained starting from Eq.~(\ref{equationofmotion}). To do this, we followed the procedure of
Ref.~\cite{raimondi2006}, namely we integrated
Eq.~(\ref{equationofmotion}) over 
$\xi = p^2/2m - \mu $ in order to obtain a Boltzmann-like equation
for the quasiclassical Green function, 
\begin{equation}
 \check g( \hat {\bf p}, {\bf x} ) = \frac{i}{ \pi } \int d \xi \, G^{R/A}({\bf p}, {\bf x})
.\end{equation}
Technically the equations simplify considerably by using that the
retarded and advanced quasiclassical Green functions are given by a
constant, $g^{R,A} \approx  \pm 1$.  
.

After some more steps we found, in the diffusive limit, when the spin splitting due to the intrinsic spin-orbit interaction, $\alpha p_F$, 
is smaller than the disorder broadening, $\tau^{-1}$ the spin current
as
\begin{equation}
\label{spincurrent}
 {j}^{a}_i = -D {\tilde \partial}_i  s^a +
 \sigma^{sH}_{\rm ext} \varepsilon_{iab} E_b + \gamma_{ij}^a  E_j
.\end{equation}
The contribution from the extrinsic spin-orbit mechanism  has the form first predicted by Dyakonov and Perel\cite{dyakonov1971} and is due
to $\sigma^{sH}_{\rm ext} = \sigma^{sH}_{ss} + \sigma^{sH}_{sj}$ with
the standard expressions for the side-jump (sj) and skew-scattering (ss) contributions
\cite{tse2006}
\begin{equation}
\label{extrinsic}
 \sigma^{sH}_{sj}  =  \sigma_D \frac{\lambda_0^2}{4}\frac{m}{e\tau},
 \hskip 0.3cm \sigma^{sH}_{ss}  = \frac{1}{4}  (p_F l) ( 2\pi N_0 v_0
 ) \sigma^{sH}_{sj},
\end{equation}
$\sigma_D=2e^2N_0 D$ being the Drude conductivity and $D=v_F^2\tau /2$ the diffusion coefficient.
It is worthwhile pointing out that the anomalous disorder-dependent
contribution to the current is one half of the side-jump contribution,
${j}_{av,i }^a=(1/2)\sigma^{sH}_{sj} \epsilon_{iab}E_b$.

The intrinsic spin-orbit coupling yields a contribution 
to the spin current
due to the $SU(2)$ magnetic field ${\bf B}^a = \tilde \partial_{\bf x}
\times {\bf A}^a$,
\begin{equation}
\gamma_{ij}^a E_j  = \frac{ \tau \sigma_D}{8me} 
({\bf B}^a \times {\bf E})_i. \end{equation}
For the Rashba model the only non-zero component of ${\bf B}^a$ is
$B^z_z= 2 (2 m \alpha)^2$ leading to
$ \gamma_{ij}^a E_j =  \varepsilon_{iab}\sigma_{\rm int}^{sH}E_b$ with
$\sigma_{\rm int}^{sH} = \frac{e}{2 \pi} (\alpha p_F \tau)^2$.
The extra terms of Ref.~\cite{lifshits2009}, associated to $\check \Sigma^1_a$,  are not relevant in the
present context and therefore have been ignored in
Eq.~(\ref{spincurrent}) 

\section{The spin Hall conductivity}
After these formal aspects on how to derive Eqs.~(\ref{continuity}) and
(\ref{spincurrent}) we now apply the equations to a specific problem,
namely the spin Hall effect\cite{dyakonov1971,hirsch1999,zhang2000,murakami2003,sinova2004}. More precisely we study the transverse
spin current that is generated by a uniform  electric field applied along the $x$-axis.  
According to Eq.(\ref{spincurrent}) the spin current flowing along the $y$-axis reads
\begin{equation}
\label{spincurrentyz}
j_y^z=  2m\alpha D s^y+(\sigma_{ext}^{sH}+\sigma^{sH}_{int})E,
\end{equation}
where the first term originates from the covariant derivative defined in Eq.(\ref{covariant}).
The presence of the in-plane spin density, $s^y$, requires that
Eq.~(\ref{spincurrentyz}) must be solved together with the
$y$-component of the continuity equation (\ref{continuity})
\begin{equation}
\label{continuitysy}
\partial_t s^y +2m\alpha \left( j^z_y-\frac{1}{2}\sigma_{sj}^{sH}E\right)
+\frac{1}{\tau_s}s^y=0.
\end{equation} 
In the absence of the extrinsic mechanism ($\lambda_0=0$), as it has been noticed previously
\cite{rashba2004,dimitrova2005,chalaev2005}, the above equation implies the vanishing of the spin current and spin Hall conductivity\cite{mishchenko2004,inoue2004,raimondi2005,khaetskii2006}. As a result the in-plane spin polarization acquires the electric-field dependent value predicted by Edelstein\cite{edelstein1990}.

\begin{figure}[t]
  \centerline{ \includegraphics[width=0.49\textwidth]{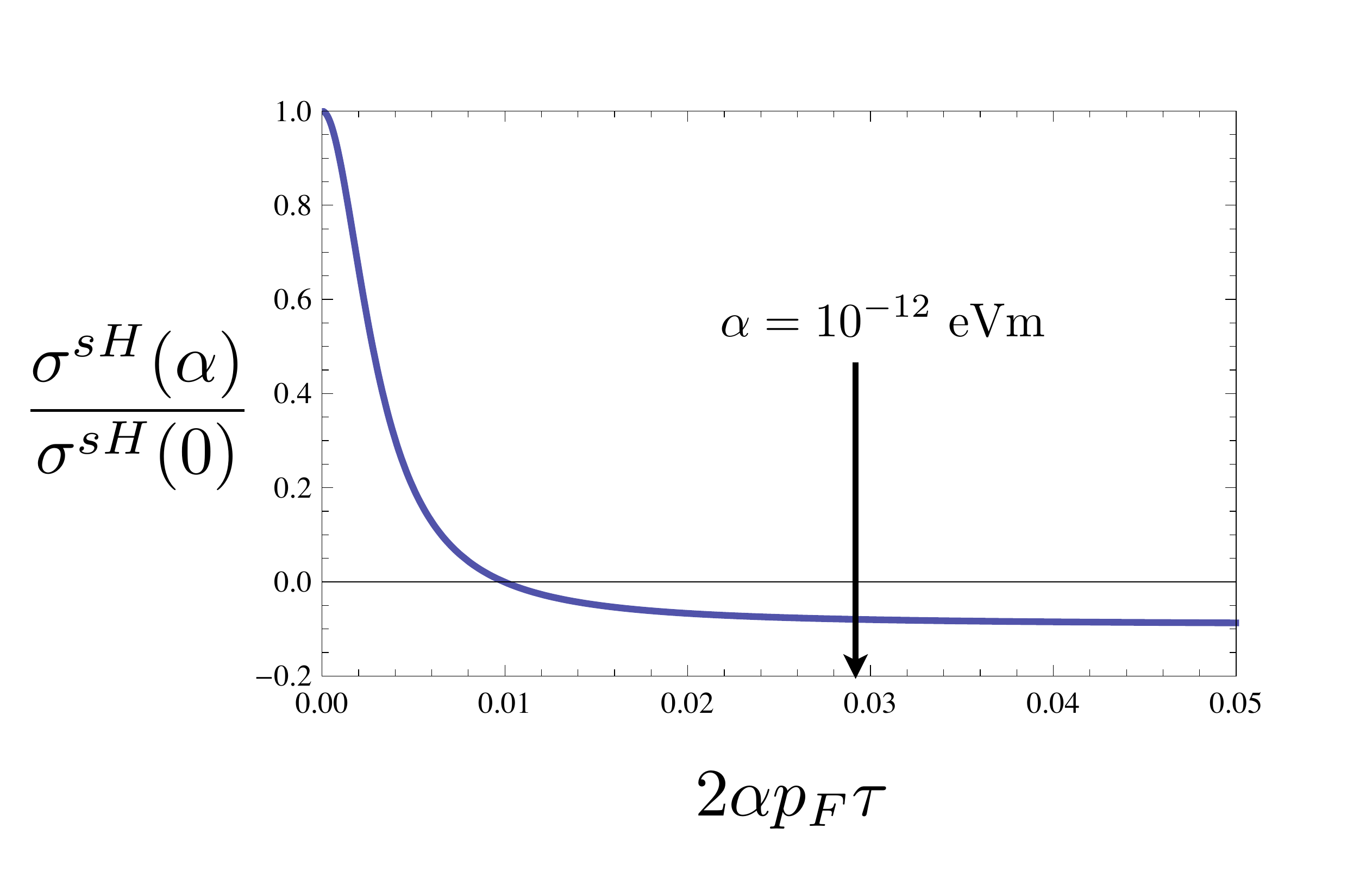} }
\caption{\label{fig2} Spin Hall conductivity as a function of the strength of the
intrinsic spin-orbit coupling. The numbers are obtained for GaAs with
density $n=10^{12}/$cm$^2$, mobility $\mu = 10^3 $ cm$^2$Vs, extrinsic
spin-orbit coupling $\lambda_0 = 4.7 \times 10^{-8}$ cm
and assuming scattering from positively charged impurities with
$N_0 v_0 = -1/2$. The spin Hall conductivity in the absence of the
Rashba coupling is $\sigma^{sH}(0)= -3 \times 10^{-7} /\Omega$. }
\end{figure}

In the presence of both the extrinsic and intrinsic mechanisms, by solving Eqs.~(\ref{spincurrentyz}) and (\ref{continuitysy}) together, one
gets $j_y^z = \sigma^{sH} E$ with
\begin{equation}\label{eq12}
 \sigma^{sH}  =  \frac{  1/\tau_s}{1/ \tau_s + 1/\tau_{DP}} \left( \sigma^{sH}_{\rm int} +
 \sigma^{sH}_{ss} + \frac{1}{2} \sigma_{sj}^{sH} \right)
 + \frac{1}{2} \sigma_{sj}^{sH}
,\end{equation}
where we have introduced the Dyakonov-Perel relaxation time $\tau_{DP}^{-1}=(2m\alpha)^2D$.
To first order in $\lambda_0^2$, $\tau_s^{-1}=0$ and one
obtains $\sigma^{sH}=(1/2)\sigma_{sj}^{sH}$ for any  $\alpha \ne 0$,
in agreement with \cite{tse2006a}. 
Clearly, the ratio $\tau_s /\tau_{DP}$ acts as a control knob of the spin Hall conductivity. 
This suggests that by adjusting the constant $\alpha$ by a suitably applied gate voltage, 
one can vary experimentally the magnitude of the spin Hall current.
For illustrative purposes in Fig. \ref{fig2}, we plot the spin Hall conductivity for GaAs
as a function of the dimensionless parameter $2\alpha p_F \tau$.

\section{Conclusions}
In summary we have presented a microscopic derivation of the equation
of motion for the Keldysh Green function in the presence of both extrinsic and intrinsic spin-orbit interaction as well as scattering from impurities. In particular we have given explicit expressions for the disorder averaged self energy at the level of Born approximation and to
the first order beyond the Born approximation. This has allowed us to derive an expression for the spin current and the associated continuity equation. 
It has been shown that the ratio of the Elliot-Yafet and Dyakonov-Perel spin relaxation times is the important parameter controlling the interplay of extrinsic and intrinsic mechanisms to the spin Hall effect in a 2DEG.

We acknowledge financial support by the DFG through SPP1285.





\begin{thebibliography}{00}
\bibitem{raimondi2009} R. Raimondi and P. Schwab,  Europhys. Lett. {\bf 87}, 37008 (2009).
\bibitem{engel2005} H.-A. Engel, B.I. Halperin, and E.I. Rashba, Phys. Rev. Lett. {\bf 95},
166605 (2005).
\bibitem{tse2006} W.-K. Tse and S. Das Sarma, Phys. Rev. Lett.  {\bf 96}, 056601 (2006). 
\bibitem{tse2006a}W.-K. Tse and S. Das Sarma, Phys. Rev. B {\bf 74}, 245309 (2006).
\bibitem{hu2006} L. Hu, Z. Huang, and S. Hu, Phys. Rev. B {\bf 73}, 235314 (2006).
\bibitem{hankiewicz2008} E.M. Hankiewvicz and G. Vignale, Phys. Rev. Lett. {\bf 100},
026602 (2008).
\bibitem{raimondi2006} R. Raimondi, C. Gorini,   P. Schwab, and M. Dzierzawa, 
Phys. Rev. B {\bf 74}, 035340 (2006).
\bibitem{rammer1986} For the connection between physical observables and Green functions see, for instance, J. Rammer and H. Smith, Rev. Mod. Phys. {\bf 58}, 323 (1986).
\bibitem{schwab2003}See for instance, P. Schwab and R. Raimondi, Annalen der Physik {\bf 12}, 471 (2003).
\bibitem{lifshits2009}M. B. Lifshits and M. I. Dyakonov, arXiv:0905.4469.
\bibitem{dyakonov1971} M. I. Dyakonov and V.I. Perel, Sov. Phys. JETP Lett. {\bf 13}, 467 (1971); Physics Letters A {\bf 35}, 459 (1971).
\bibitem{hirsch1999} J. E. Hirsch, Phys. Rev. Lett. {\bf 83}, 1834 (1999).
\bibitem{zhang2000} S. Zhang, Phys. Rev. Lett. {\bf 85}, 393 (2000).
\bibitem{murakami2003} S. Murakami, N. Nagaosa, and S.-C. Zhang, Science {\bf 301}, 1348 (2003).
\bibitem{sinova2004} J. Sinova, D. Culcer, Q. Niu, N. Sinitsyn, T. Jungwirth, and A. MacDonald, Phys. Rev. Lett. {\bf 92}, 126603 (2004).
\bibitem{rashba2004} E. I. Rashba, Phys. Rev. B {\bf 70}, 201309(R) (2004).
\bibitem{dimitrova2005} O. V. Dimitrova, Phys. Rev. B {\bf 71}, 245327 (2005).
\bibitem{chalaev2005} O. Chalaev and D. Loss, Phys. Rev. B {\bf 71}, 245318 (2005).
\bibitem{mishchenko2004} E. G. Mishchenko, A.V. Shytov, and B.I. Halperin, Phys. Rev. Lett. {\bf 93}, 226602 (2004).
\bibitem{inoue2004} J. I. Inoue, G.E.W. Bauer, and L.W. Molenkamp,  Phys. Rev. B {\bf 70}, 041303 (R) (2004).
\bibitem{raimondi2005} R. Raimondi and P. Schwab, Phys. Rev. B {\bf 71}, 033311 (2005).
\bibitem{khaetskii2006} A. Khaetskii, Phys. Rev. Lett. {\bf 96}, 056602 (2006).
\bibitem{edelstein1990} V. M. Edelstein, Solid State Commun. {\bf 73}, 233 (1990).
\end{thebibliography}
\end{document}